# Efficient Electro-optical Tuning of Optical Frequency Microcomb on a Monolithically Integrated High-Q Lithium Niobate Microdisk

Zhiwei Fang [1,2,6], Haipeng Luo [3,6], Jintian Lin [4], Min Wang [1,2], Jianhao Zhang [4], Rongbo Wu [4], Junxia Zhou [1,2], Wei Chu [4], Tao Lu [3,*], and Ya Cheng [1,2,4,5,†]

[1] *State Key Laboratory of Precision Spectroscopy, East China Normal University, Shanghai 200062, China.*

[2] *XXL—The Extreme Optoelectromechanics Laboratory, School of Physics and Electronics Science, East China Normal University, Shanghai 200241, China.*

[3] *Department of Electrical and Computer Engineering, University of Victoria, Victoria, BC, V8P 5C2, Canada.*

[4] *State Key Laboratory of High Field Laser Physics, Shanghai Institute of Optics and Fine Mechanics, Chinese Academy of Sciences, Shanghai 201800, China.*

[5] *Collaborative Innovation Center of Extreme Optics, Shanxi University, Taiyuan 030006, Shanxi, China*

[6] *Zhiwei Fang and Haipeng Luo contributed equally to this work.*

[*] *taolu@ece.uvic.ca*

[†] *ya.cheng@siom.ac.cn*

**Abstract**

We demonstrate efficient tuning of a monolithically integrated lithium niobate microdisk (LN) optical frequency microcomb. Utilizing the high optical quality (Q) factor (i.e., Q~7.1×$10^6$) of the microdisk, the microcomb spans over a spectral bandwidth of ~200 nm at a pump power as low as 20.4 mW. Combining the large eletro-optic coefficient of LN and optimum design of the geometry of microelectrodes, we demonstrate electro-optical tuning of the comb with a spectral range of 400 pm and a tuning efficiency of ~38 pm/100V.

Optical frequency comb is revolutionizing precision spectroscopy and frequency metrology [1, 2]. Miniaturization of optical frequency combs enables higher performance and lower cost for a wide range of innovative applications including optical atomic clocks [3], optical communications [4], spectroscopy [5], and quantum technology [6]. Optical microresonators based on Kerr nonlinearity materials provide an efficient approach for comb generation owing to the greatly enhanced light fields in the microresonators [7]. Up to date, optical frequency combs have been successfully demonstrated in silica [8], semiconductor [9-11], and crystalline whispering gallery mode (WGM) resonators [12-14]. Both thermal and electric tuning of the frequency combs have been achieved by exploiting the thermo-optic and electro-optic effects of various types of microresonator materials [15-17]. For fast frequency switching and stabilization as well as wide-range frequency tuning, it is highly desirable to enhance the tuning efficiency, tuning range as well as tuning rate. Here, we demonstrate efficient and fast tuning of an optical frequency comb taking the advantage of the high electro-optic coefficient of lithium niobate (LN). The microdisk fabricated on lithium niobate on insulator (LNOI) reaches an optical quality (Q) factor as high as $7.1 \times 10^6$, allowing us to generate a comb with a bandwidth of ~200 nm at a pump power as low as 20.4 mW. In addition, optimization of geometry of the microelectrodes results in a tuning efficiency of ~38 pm/100V and a tuning range of ~400 pm as observed in our experiment.

Our monolithically integrated comb is fabricated using chemo-mechanical polish,

which is detailed in Supplementary Material S1. Briefly speaking, a thin layer of chromium (Cr) film is first deposited on the surface of the LNOI by magnetron sputtering to serve as a mask layer. The Cr film will then be etched to a circular pattern. In the subsequent chemo-mechanical polish, the LN film underneath the Cr mask can be preserved whereas the LN in the opening area is completely removed. This step allows us to create a LN microdisk with extremely smooth rim for high Q factors [18]. The microelectrodes are fabricated directly by patterning the Cr film using femtosecond laser writing [19]. Unlike the other laser fabrication technologies that utilize long-pulsed lasers (i.e., nanosecond and picosecond pulsed lasers), the thermal effects in the femtosecond laser micromachining can be significantly suppressed to enable the high fabrication resolution of ~500 nm [20-22].

Figure 1(a) presents the top view of the 100-μm-diameter LN microdisk scanning electron micrograph (SEM). The silica pedestal underneath the microdisk has a diameter of ~60 μm. As indicated in Fig. 1, the anode is fabricated into a circular pad of a comparable diameter to overlap the area supported by the silica pedestal while the cathode has a concave semicircle pattern with a diameter of ~130 μm surrounding the LN microdisk. The Cr microelectrodes are clearly visible in the optical micrograph under transmission illumination in Fig. 1(b), which appear completely dark in contrast to the transparent LN microdisk. The simulation of the static electric potential distribution on LN disk is given in the Supplementary Material S2.

The sideview SEM reveals that the wedge angle is 12° and confirms that the LN microdisk has a thickness of 800 nm. The monolithically integrated microcomb is fabricated with a one-step seamless femtosecond laser writing which simultaneously defines the LN microdisk and the microelectrodes with a positioning precision of ~100 nm. This one-step seamless writing strategy mitigates the misalignment between the microelectrodes and the LN microdisk, thus the positioning precision limit is dominantly determined by the motion stage precision.

To characterize the tunability of the microcomb, we used an experimental setup as shown in Fig. 2(a). Here, a continuous-wave telecom-band tunable laser (TLB 6728, New Focus Inc.) was used as the pump source. The tunable laser, of which the polarization state can be adjusted by an in-line fiber polarization controller, was amplified by an erbium-doped fiber amplifier (EDFA, Shenzhen Golight Technology Co. Ltd.). The noise of EDFA was suppressed by a tunable bandpass optical fiber filter. A tapered fiber with a waist of 1 μm was used to couple the light into and out of the fabricated LN microdisk. 90% of the output beam was directed to an optical spectrum analyzer (OSA: AQ6370D, YOKOGAWA Inc.) for spectral analysis using a fiber beam splitter, whereas the remaining 10% of the output beam was routed to a power meter (PM100D, Thorlabs Inc.) for input pump power monitoring. A direct current (DC) stabilized power source (CE1500002T, Rainworm Co. Ltd.) was used as the voltage generator for Cr electrodes, which provided a variable voltage ranging from 0 V to 500 V. Two probes (ST-20-0.5, GGB Industries Inc.) to apply DC voltage on Cr electrodes respectively.

Fig. 2(b) illustrates the measured transmission spectrum near the wavelength of 1552 nm at a low pump power where the thermal broadening effect is negligible. The Q factors of two split modes are identified to be $5.3 \times 10^6$ and $7.1 \times 10^6$ through a double Lorentzian fitting. A Raman-assistant frequency comb spectrally centered at 1552.4 nm was observed by increasing the pump laser power to 20.4 mW as shown in Fig. 2(c). The spectral range of comb is above 200 nm. We observed that the pump power of the comb was significantly lower than the previously reported owing to the higher Q factors of the LNOI microdisk fabricated by chemo-mechanical polish [14, 23].

Fig. 3 shows the evolution of the comb recorded at a wavelength resolution of 0.4 nm with the increasing pump wavelength. As shown in Fig. 3(a), the Stokes and anti-Stokes Raman processes occur at a pump wavelength of 1552.39 nm. When the pump wavelength is red-tuned to 1552.68 nm, Raman-assisted four-wave mixing generation appears near the pump, Stokes and anti-Stokes wavelengths as shown in Fig. 3(b). By further detuning the pump laser to 1552.72 nm and 1552.74 nm wavelengths, a comb with a line spacing of double free-spectral range (FSR: 3.6 nm) was generated as shown in Fig.3 (c) as well as a comb with a line spacing of single FSR as shown in Fig.3 (d), respectively. It is noteworthy that our results are consistent with the Raman-assisted frequency comb generated in various types of optical microresonators as reported recently [24, 25]. In general, the Raman-assisted frequency comb generation involves both Stimulated Raman scattering (SRS) and Kerr nonlinearity. While the bandwidth of Kerr frequency comb is mainly limited by cavity dispersion, the SRS in microresonators are governed by the frequency matching between the cavity resonances and the Raman gain, which helps expand the comb bandwidth towards both red and blue by the Stokes or anti-Stokes processes, respectively.

We demonstrated the real-time tuning of the comb by adding a tunable electric voltage on the Cr microelectrodes. We observe that by increasing the electric voltage from ~-100 V to ~300 V, we are able to continuously red-shift the resonant wavelength by ~150 pm as shown in Fig. 4(a). Here, the measurement was performed around the resonant wavelength of 1552.4 nm. The fitting in Fig 4(b) confirms that the wavelengths of comb lines linearly change with the increasing electric voltage at a tuning rate of ~38 pm/100 V. As further revealed in Fig. 4(c), a tuning range of ~400 pm can be obtained by varying the electric voltage between -500 V and 500 V. The tuning range covers one ninth of the FSR of our LNOI microdisk of a diameter of 100 μm. At this moment, the tuning range is limited by the electric breakdown of LN at higher electric voltages.

In summary, electro-optical tuning of a monolithically integrated frequency comb based on an LNOI microresonator has been demonstrated, which offers great potential for the relevant applications. Further research will be carried out on increasing the tuning range by systematically improving the design of the LNOI microdisk and that of the microelectrodes.

The work is supported by the National Natural Science Foundation of China (Grant Nos. 11734009, 11874375, 61590934, 11874154, 61505231, 11604351, 11674340, 61575211, 61675220, 61761136006), NSERC Discovery (RGPIN2015-06515), the Strategic Priority Research Program of Chinese Academy of Sciences (Grant No. XDB16030300), the Key Research Program of Frontier Sciences, Chinese Academy of Sciences (Grant No. QYZDJ-SSW-SLH010), Key Project of the Shanghai Science and Technology Committee (Grant No. 17JC1400400, 18DZ1112700). T. L. and H. L. would like to acknowledge CMC Microsystems.

**Figure captions:**

Fig. 1 (a) Top view SEM of the 100-μm LN microdisk integrated with Cr electrodes and (b) optical micrograph of the LN disk and integrated Cr electrodes under transmission illumination. (c) Side view SEM of the fabricated LN microdisk integrated with Cr electrodes, Inset (upper right) shows the LN microdisk have a thickness of 800 nm and wedge angle of 12°.

Fig.2 (a) Schematic of the experimental setup for tunable comb characterization in the integrated comb device. (EDFA: erbium-doped fiber amplifier; PD: photodetector; OSA: optical spectrum analyzer. (b)The double Lorentzian fitting showing a mode splitting, indicating the Q-factors of $5.3\times10^6$ and $7.1\times10^6$ as measured at 1552.4 nm wavelength. (b) Optical spectra of comb generation.

Fig. 3 Raman-assisted frequency comb generation, recorded with a resolution of 0.4 nm by OSA. (a) Stokes and anti-Stokes Raman scatterings before comb generation. (b) Raman-assisted four-wave mixing generation. (c) Comb generation with space of double FSR. (d) Comb generation with space of single FSR.

Fig. 4 (a) The resonance continuously red-shifted shifts by increasing the applied voltage. (b) The resonant wavelength shifts linearly with an electrical tuning efficiency of ~38 pm/100V.; (c) The comb line shift as a function of the applied voltage, recorded with a resolution of 0.1 nm by OSA.

Figure 1

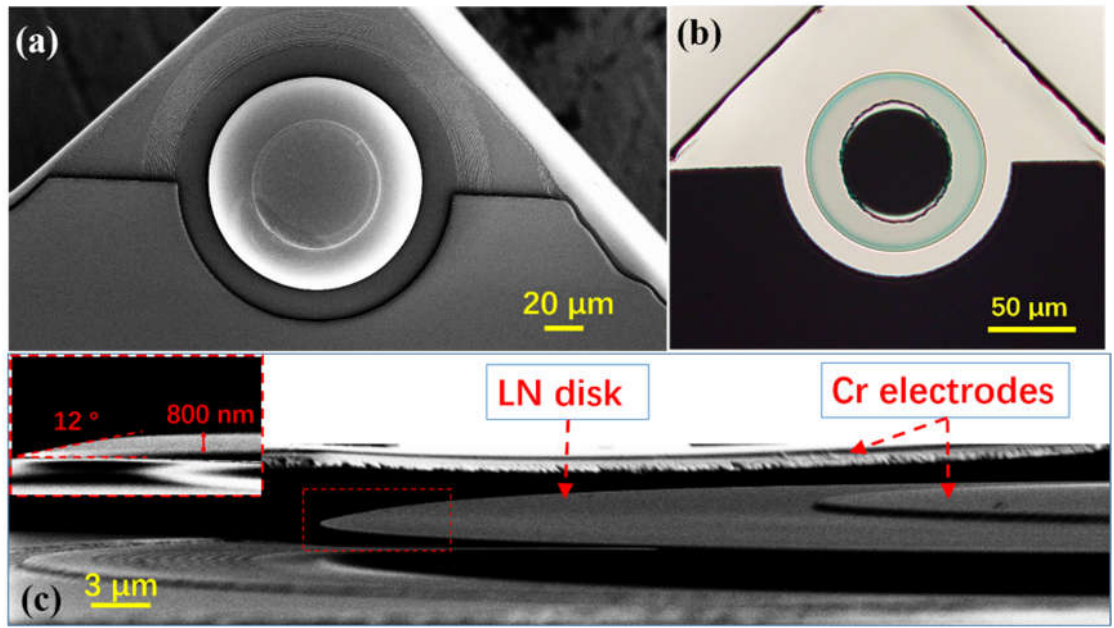

Figure 2

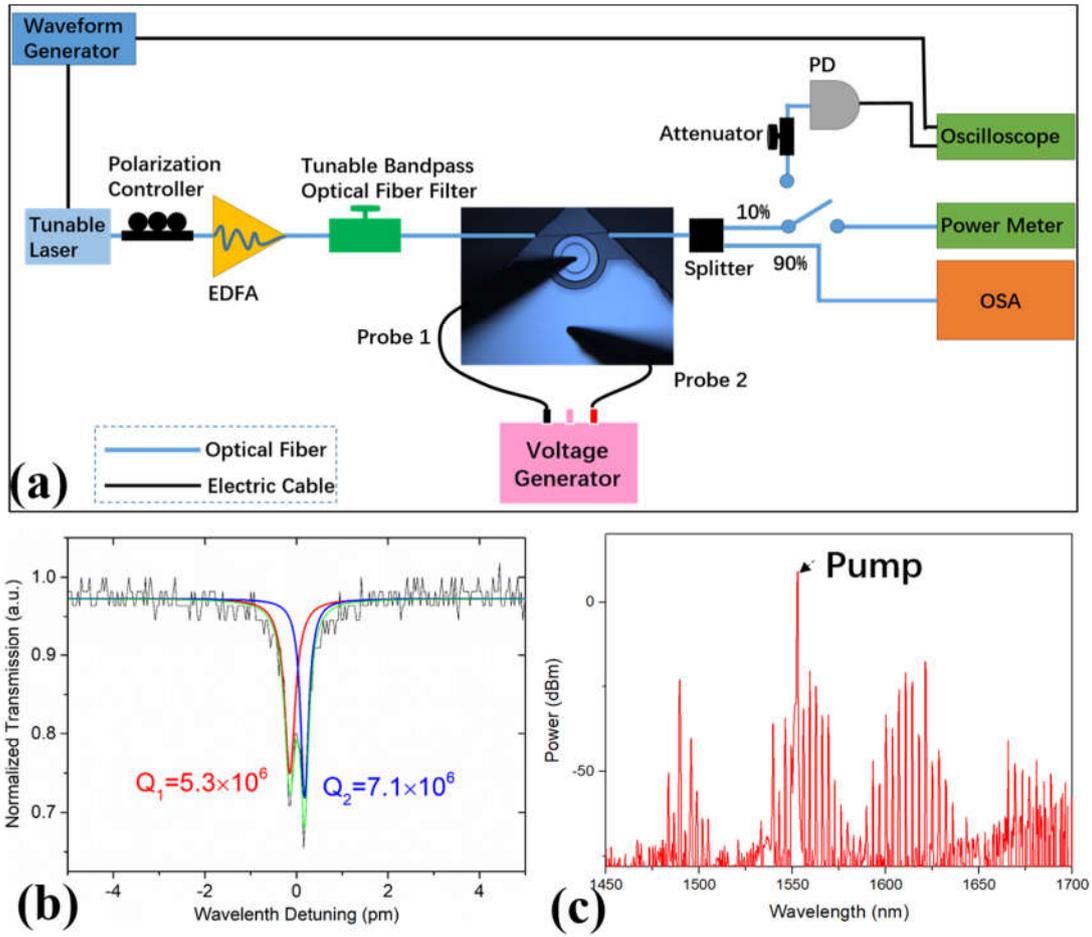

Figure 3

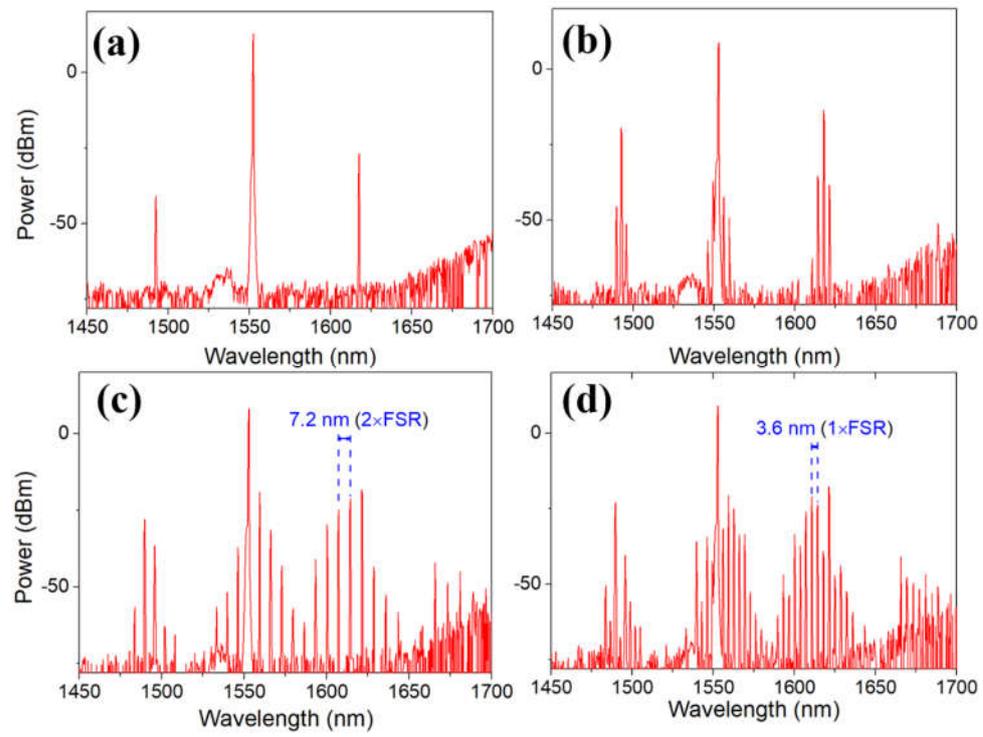

Figure 4

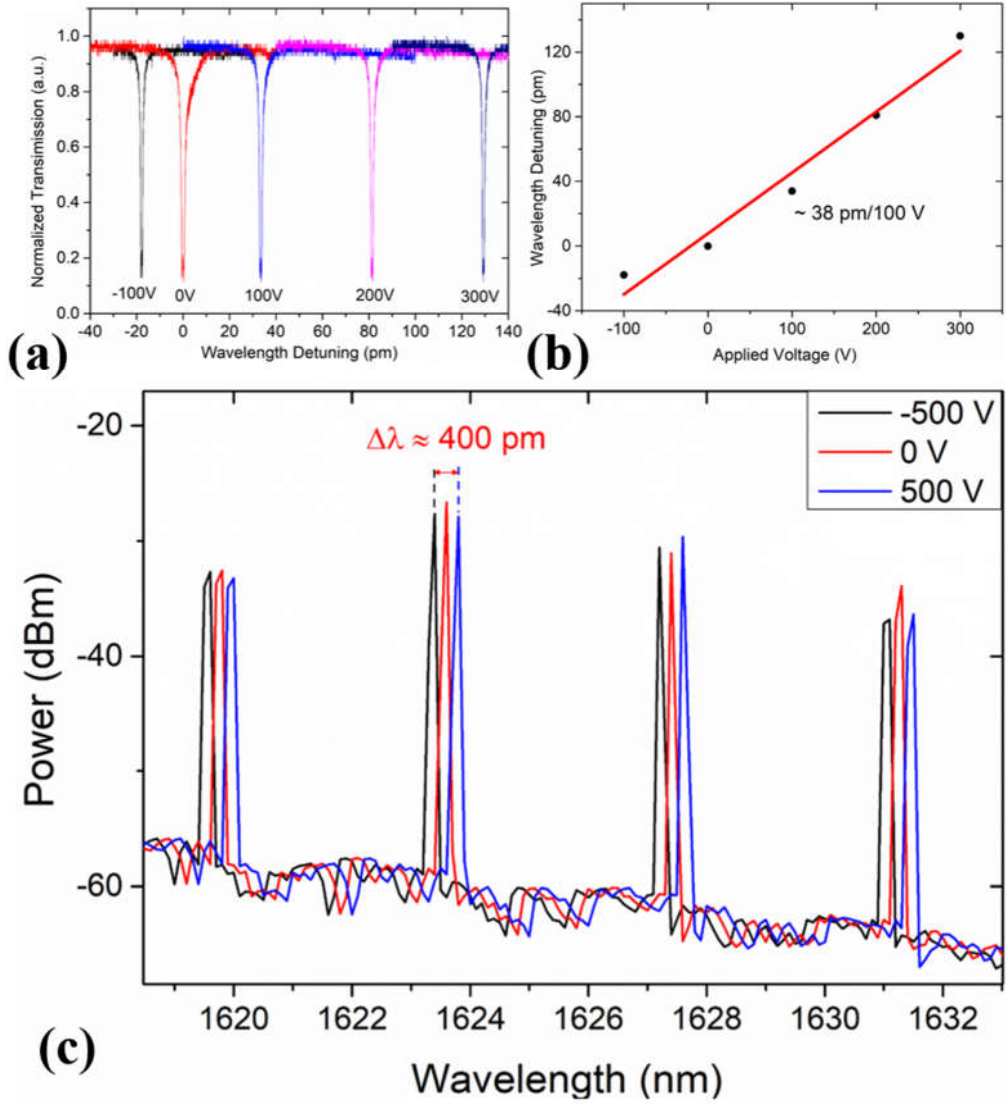

# Supplementary materials

# Efficient Electro-optical Tuning of Optical Frequency Microcomb on a Monolithically Integrated High-Q Lithium Niobate Microdisk

Zhiwei Fang [1,2,6], Haipeng Luo [3,6], Jintian Lin [4], Min Wang [1,2], Jianhao Zhang [4], Rongbo Wu [4], Junxia Zhou [1,2], Wei Chu [4], Tao Lu [3,*], and Ya Cheng [1,2,4,5,†]

[1] State Key Laboratory of Precision Spectroscopy, East China Normal University, Shanghai 200062, China.

[2] XXL—The Extreme Optoelectromechanics Laboratory, School of Physics and Electronics Science, East China Normal University, Shanghai 200241, China.

[3] Department of Electrical and Computer Engineering, University of Victoria, Victoria, BC, V8P 5C2, Canada.

[4] State Key Laboratory of High Field Laser Physics, Shanghai Institute of Optics and Fine Mechanics, Chinese Academy of Sciences, Shanghai 201800, China.

[5] Collaborative Innovation Center of Extreme Optics, Shanxi University, Taiyuan 030006, Shanxi, China

[6] Zhiwei Fang and Haipeng Luo contributed equally to this work.

* *taolu@ece.uvic.ca*

† *ya.cheng@siom.ac.cn*

# S 1 Details of device fabrication

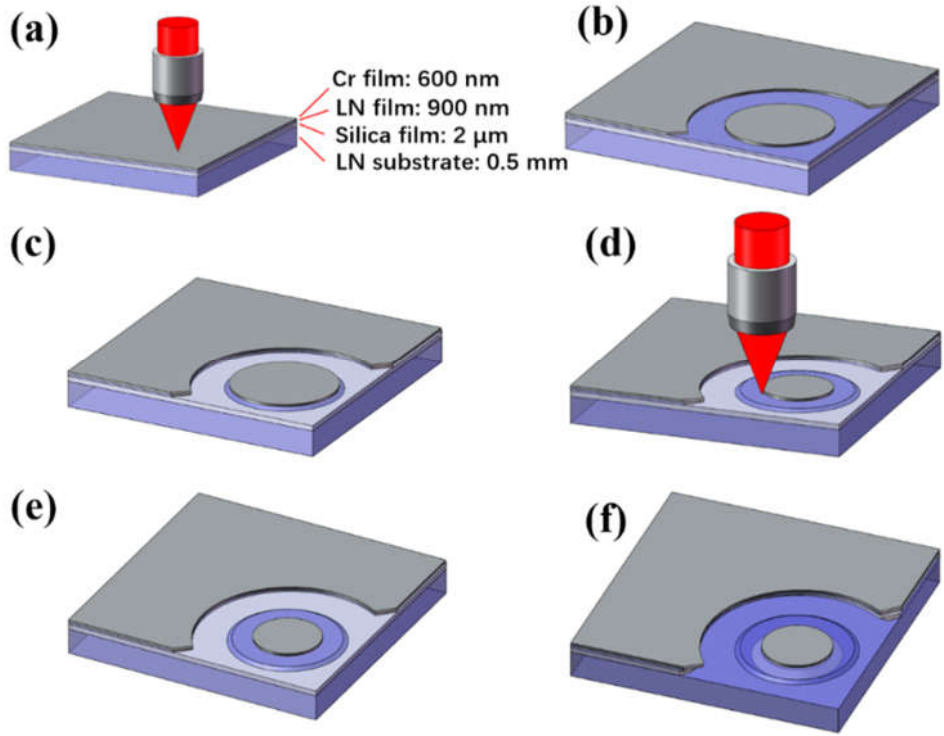

Fig. S1 The processing flow of fabricating an on-chip LN microdisk integrated with electrodes is illustrated. (a)-(b) Patterning the Cr thin film into a circular mask and the circular cathode using femtosecond laser microfabrication. (c) Transferring the circular disk-shaped mask to LNOI by chemo-mechanical polishing. (d) Selective removal of the disk-shaped Cr mask to form the anode on the LN microdisk using femtosecond laser ablation. (e) Reducing the thickness of LN disk by a secondary chemo-mechanical polishing. (f) Chemical wet etching of the sample to undercut the fused silica beneath the LN microdisk, forming the freestanding LN microdisk resonator supported by the fused silica pedestal.

In our experiment, the on-chip LN microdisk resonator integrated with Cr electrodes was fabricated on a commercially available X-cut LN thin film wafer with a thickness of 900 nm (NANOLN, Jinan Jingzheng Electronics Co. Ltd.). The LN thin film is bonded by a silica layer with a thickness of ~2 μm on a 0.5 mm thick LN substrate, and a 600-nm-thickness layer of chromium (Cr) film was deposited on the surface of the LNOI by magnetron sputtering method.

The fabrication process includes five steps, as illustrated in Fig. 1. First, the Cr film on the LNOI sample was patterned into a disk-shaped mask (circular) and the cathode of a semicircular shape surrounding the LN microdisk using space-selective femtosecond laser (PHAROS, LIGHT CONVERSION Inc.). Subsequently, the chemo-mechanical polishing (CMP) process was performed to fabricate the LN microdisk by a wafer polishing machine (NUIPOL802, Kejing Inc.). Next, the femtosecond laser ablation was carried out again to reduce the diameter of the Cr mask to form the anode. Then, a secondary CMP process was performed for thinning the LN disk as well as cleaning the ablation debris on the disk surface. Finally, the fabricated structure was immersed in a buffered hydrofluoric acid (HF) solution to partially under etch the silica layer into the shape of a pillar supporting the LN microdisk. It takes about 1 hour in total to produce the whole integrated device. More details of the fabrication process can be found in Refs. [S1-S3].

## S2 Numerical simulation of the electric field in the LN microdisk

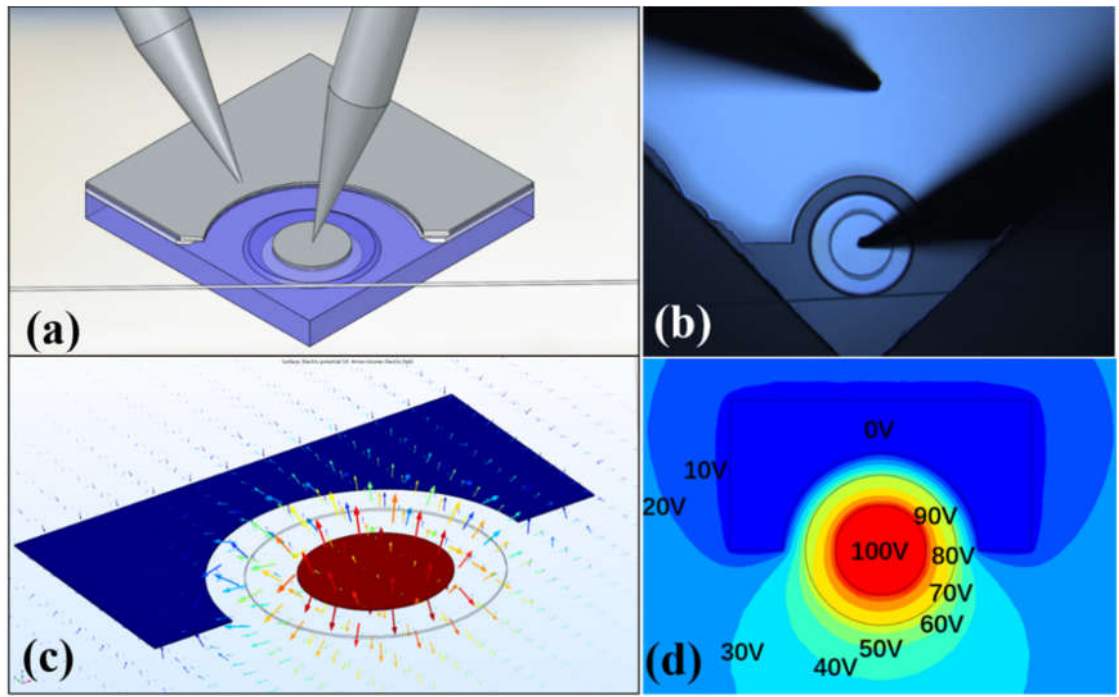

Fig. S2 (a) Schematic of the integrated comb device. (b) Optical micrograph showing the top view of the integrated comb device (LN microdisk and Cr electrodes), the tapered fiber, and the probes. (c) The simulated distribution of the electric field in integrated comb device. (d) Contours of electric potential at XY plane of integrated comb device when an 100 V voltage was applied to the Cr electrodes.

Figure S2 (a) illustrates the schematic of the integrated comb device. The comb is generated by coupling a pump laser using a fiber taper. The microelectrodes are activated with a pair of microneedles for realizing tuning of the comb. Figure S2(b) presents the optical micrograph of the integrated comb device which is under the characterizations of comb generation and comb tunability. The simulated distribution of the electric field in integrated comb device is shown in Fig. S2(c), showing that the field is of the highest strength in the LN microdisk and points outward from the center of the microdisk. This feature is also confirmed by the illustrated contours of electric potential at XY plane of integrated comb device as shown in Fig. S2(d).